# Room-temperature alignment-free magnetometry with boron vacancies in hot-pressed hexagonal boron nitride

Shuyu Wen,[1,2,3*] Raul Coto[4*], Peiting Wen,[1,5] Slawomir Prucnal,[1] Manfred Helm,[1,5] Jun-Wei Luo,[2,3] Shengqiang Zhou,[1] and Yonder Berencén[1*]

[1]Helmholtz-Zentrum Dresden-Rossendorf, Institute of Ion Beam Physics and Materials Research, Bautzner Landstrasse 400, 01328 Dresden, Germany

[2]State Key Laboratory of Semiconductor Physics and Chip Technologies, Institute of Semiconductors, Chinese Academy of Sciences, Beijing 100083, China

[3]Center of Materials Science and Optoelectronics Engineering, University of Chinese Academy of Sciences, Beijing 100049, China

[4]Department of Chemistry and Physics, Nova Southeastern University, Fort Lauderdale-Davie, Florida 33328-2004, USA

[5]Technische Universität Dresden, 01062 Dresden, Germany

Magnetic field sensing is essential for applications in communication, environmental monitoring, and biomedical diagnostics. Quantum sensors based on solid-state spin defects, such as nitrogen-vacancy centers in diamond or boron vacancies in single-crystal hexagonal boron nitride (hBN), typically require precise alignment between the external magnetic field and the defect's spin quantization axis to achieve reliable sensing. This alignment constraint complicates device integration and hinders scalability. Here, we demonstrate room-temperature optically detected magnetic resonance (ODMR) from negatively charged boron vacancies ($V_B^-$) in commercially available hot-pressed polycrystalline hBN. The random grain orientation inherently samples a broad range of spin quantization axes, enabling alignment-free magnetic field detection. Numerical modeling further confirms that sensing remains feasible despite anisotropic sensitivity, establishing hot-pressed hBN as a robust and practical platform for quantum magnetometry. This approach paves the way toward low-cost, scalable, and mechanically stable quantum magnetic field sensors suitable for real-world deployment.

*Authors to whom correspondence should be addressed: s.wen@hzdr.de, rcotocab@nova.edu, y.berencen@hzdr.de

**Introduction**

Magnetic field sensing plays a key role in a wide range of applications spanning communication[1,2], medical diagnostics[3] and environmental monitoring[4]. Recently, solid-state spin defects in materials such as hexagonal boron nitride (hBN) have emerged as promising candidates for highly sensitive and compact magnetic field sensors[5]. Among these, the negatively charged boron vacancy (VB$^-$) defect, with a spin state S=1, has attracted significant interest due to its stable optical transitions and magnetic field sensitivity at room temperature[6,7].

Optically detected magnetic resonance (ODMR) serves as a powerful technique to probe such spin defects, enabling their application as quantum sensors[6]. To date, most studies have focused on magnetic field sensing using single-crystal hBN flakes, where the defect spin quantization axis is well-defined and often aligned with the external magnetic field[8,9]. However, a fundamental challenge in using S=1 spin systems like VB$^-$ for magnetometry lies in their intrinsic anisotropy: the spin transitions are primarily sensitive to the component of the magnetic field projected along the spin quantization axis[10,11]. Misalignment between the external field and this axis leads to reduced ODMR contrast and degraded sensitivity[12]. This alignment dependence similarly constrains sensing performance in other platforms, such as nitrogen-vacancy (NV) centers in diamond[10].

In this work, we investigate commercially available hot-pressed polycrystalline hBN substrates hosting VB$^-$ centers created via helium irradiation as a platform for room-temperature magnetic field sensing that is operationally alignment-free. The key advantage of hot-pressed hBN lies in its polycrystalline structure, where the random, or mildly textured, grain orientations naturally sample a broad range of spin quantization axes. This enables magnetic field detection without the need for precise alignment of the sensor relative to the external field, addressing a key limitation of single-crystal hBN flakes and NV centers in diamond. While the sensitivity remains direction-dependent due to the anisotropic distribution of defect orientations, our numerical simulations show that vector magnetic field sensing is feasible through calibration. We support this with experimental ODMR measurements under different orientations of an external magnetic field, demonstrating the functionality of the sensor and validating the modeling approach. Combined with its scalability, mechanical robustness, and compatibility with large-area fabrication, hot-pressed hBN emerges as a promising platform for practical, alignment-free quantum magnetic sensors suitable for real-world applications.

**Results and Discussion**

We created negatively charged boron vacancies in commercially available hot-pressed hBN samples by locally irradiating them with He$^+$ ions at a fluence of $2\times10^{16}$ cm$^{-2}$ with an energy of 1.7 MeV[7]. This process created an irradiated region on the hBN substrate, which exhibits a reddish-brown color (Figure. 1a), indicating the generation of color centers under these He irradiation conditions[13]. Subsequently, we conducted Raman and photoluminescence (PL) spectroscopy experiments to characterize the optical properties of the irradiated hBN samples.

Figure. 1b depicts a sharp Raman peak at 1366 cm$^{-1}$ observed in both the reference pristine hBN and He$^+$-irradiated hBN samples. This peak corresponds to the intralayer E$_{2g}$ vibration mode in hBN, originating from in-plane out-of-phase vibrations of adjacent boron and nitrogen atoms in the same layer[13]. Following He$^+$ ion irradiation, a V$_B^-$ defect-related vibration mode emerges distinctly at 1295 cm$^{-1}$, indicating the creation of V$_B^-$ in the hBN sample[7].

The PL measurement results of the hBN samples are presented in Figure. 1c. The emission peak near 800 nm is a clear fingerprint of the PL from the V$_B^-$ defect in hBN generated upon He$^+$ irradiation[7], which was not observed before irradiation. Additionally, the broad PL emission peak near 620 nm is attributed to the emissions from both boron vacancies and substitutional boron defect[14].

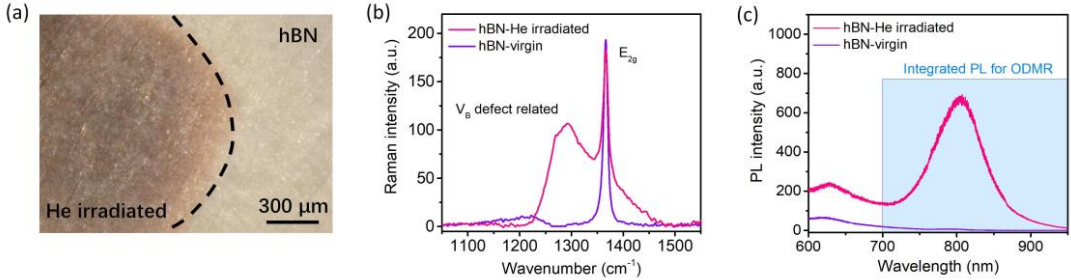

**Figure 1. Optical characterization of the He-irradiated hot-pressed hBN. a,** Optical microscope image showing the commercially available hot-pressed hBN sample used in this study. The dashed line delineates a part of the locally irradiated region on the hBN substrate, characterized by a reddish-brown color indicative of color center formation. **b,** Raman spectrum and **c,** Photoluminescence spectrum of hBN samples before and after He$^+$ irradiation. The light blue area indicates the entire spectral region in which the PL intensity was integrated for ODMR measurements.

The color center V$_B^-$ in hBN material represents a type of spin defect characterized by the absence of a boron atom within the hBN lattice, resulting in a negative charge. This defect exhibits behavior akin to artificial atoms, featuring efficient optical transitions[15]. The V$_B^-$ defect center possesses a spin quantum number of S=1, which leads to the splitting of the triplet ground state (GS) into m$_s$=0 and m$_s$=±1 states, as illustrated in Figure. 2a. Notably, the energy gap associated with this GS splitting falls within the microwave energy range, enabling resonant pumping through microwave excitation.

To investigate the ODMR of V$_B^-$ in hot-pressed hBN, we positioned the sample above a printed circuit board (PCB) holder with a 1-mm-wide Au strip-line microwave waveguide, where the generated microwave field induces rotations of the V$_B^-$ spin GS, allowing for readout via PL (Figure. 2b). Utilizing a home-built measurement setup based on signal synchronization and lock-in signal acquisition techniques (Figure. 2c), we conducted ODMR measurements on He$^+$-irradiated hBN substrates. This involves sweeping the frequency of the microwave field while simultaneously optically pumping with a 532 nm laser and measuring the PL intensity using a silicon avalanche photodetector. When the microwave frequency is not resonant with spin transitions, the defect remains spin-polarized in the m$_s$=0 state, resulting in maximum PL intensity.

However, upon resonance, the spin is rotated toward $m_s=\pm 1$, leading to a reduction in PL intensity due to enhanced non-radiative relaxation through intersystem crossing (ISC) assisted by a singlet metastable state (MS). This latter non-radiative mechanism is favored from the $m_s=\pm 1$ spin sublevels of the excited state (ES) to the $m_s=0$ spin sublevel of the GS.

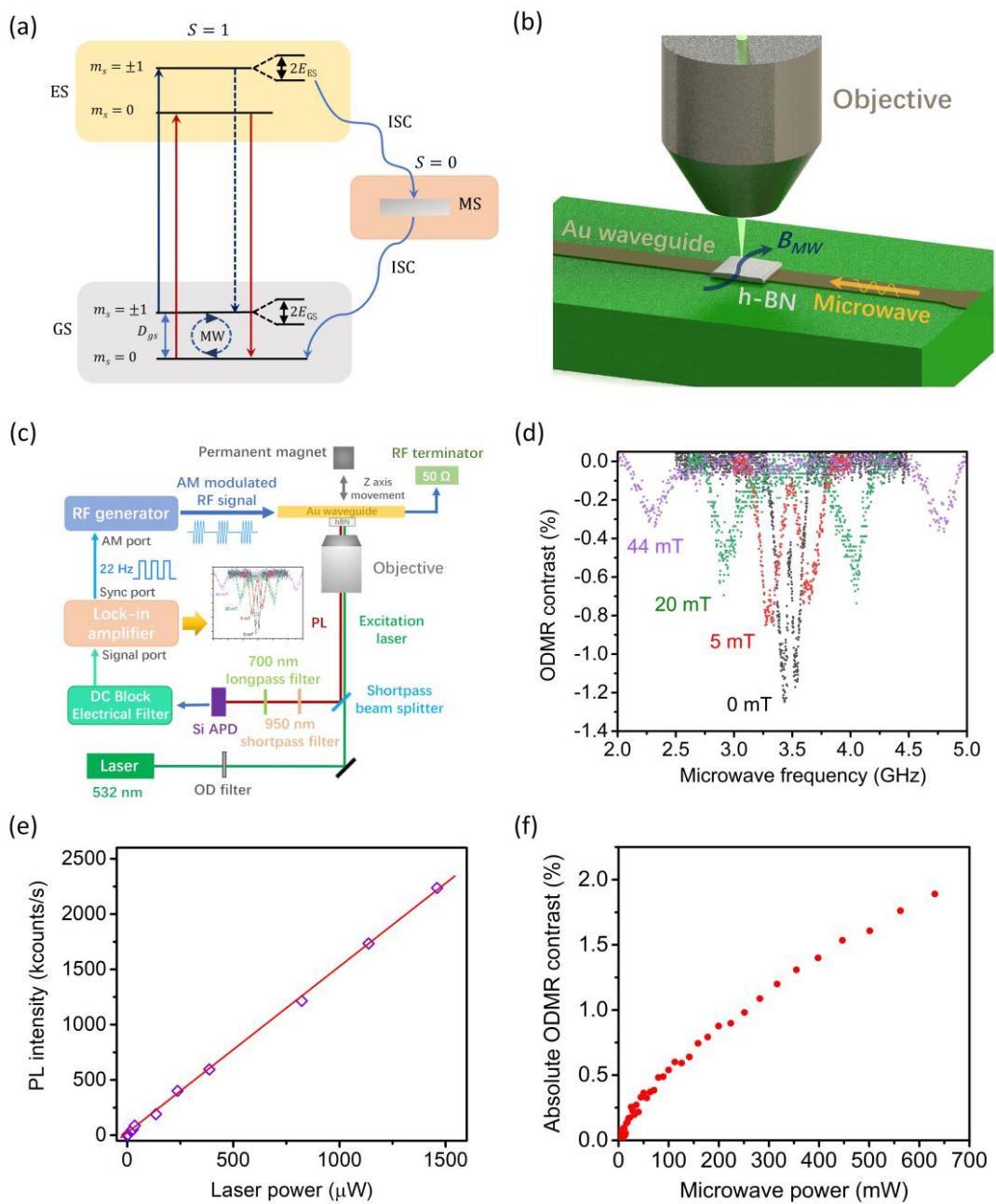

**Figure 2. Energy level structure, experimental setup, and ODMR measurements of VB⁻ in hBN. a,** Energy level diagram illustrating the optical transitions of the defect and its spin sublevels. In both the ground state (GS) and the excited state (ES), the spin triplet levels are split by $D_{GS}$ and $D_{ES}$, respectively. A non-radiative intersystem crossing (ISC) preferentially occurs from the $m_s=\pm 1$ spin sublevels of the ES to a metastable (MS) state, from which the system decays to the GS. The dashed line represents a spin-dependent optical transition with less contribution to the photoluminescence process. **b,** Schematic

representation of the setup for probing ODMR of $V_B^-$ in hot-pressed hBN. The sample is positioned above a coplanar waveguide, and the generated microwave field induces rotations of the $V_B^-$ spin ground state, allowing for readout via photoluminescence. **c,** Schematic of the home-built experimental setup. **d,** ODMR contrast as a function of driving microwave frequency at various external magnetic fields aligned perpendicular to the sample surface. **e,** Photoluminescence (PL) of $V_B^-$ in hBN sample under different laser excitation powers. **f,** ZFS ODMR contrast of $V_B^-$ as a function of different microwave powers at a frequency of 3420 MHz.

Thus, we conducted microwave frequency sweeping from 2.0 to 5.0 GHz under different magnetic fields for room-temperature ODMR experiments, and the results are depicted in Figure. 2d. To enhance the ODMR contrast, defined as $\Delta PL/PL = (PL_{off\text{-}resonance} - PL_{on\text{-}resonance})/PL_{off\text{-}resonance}$, we integrated the PL intensity within the spectral window from 700 nm to 950 nm, highlighted as the blue region in the PL spectrum of Figure.1c.

At first and upon these experimental conditions, we performed ODMR measurements with no external magnetic fields, resulting in an ODMR resonant peak centered at 3.48 GHz, indicative of a longitudinal zero-field splitting (ZFS) parameter $v_0 = D/h = 3.48$ GHz[5,15] (Figure. 2d). This main ODMR resonant peak is composed of two peaks located at 3.42 GHz and 3.54 GHz under a zero external magnetic field, known as E-splitting between GS $m_s = +1$ and $m_s = -1$ (Figure. 2a). This splitting is attributed to the interaction between the $V_B^-$ electron spin and a local electric field in hBN crystal[16]. We observed a splitting indicative of the transverse ZFS parameter $E/h = 60$ MHz, which is similar to previous reports of $V_B^-$ in hBN flakes (viz. 50-126 MHz) [8,17]. The $E$ parameter is expected to vary depending on local strain caused by crystal quality, sample homogeneity, and different sample mounting techniques[13], rather than sensitivity from thermal effects, which may arise due to, for instance, fluctuations in laser excitation power[18]. The relatively low $E$ factor observed in our bulk hBN sample is thus attributed to the bulk properties of the polycrystalline hBN material, which keeps the $V_B^-$ center relatively isolated from external environmental effects.

We observed that under different external magnetic fields ($B$) applied along the out-of-plane direction, the ZFS peaks are further split due to the Zeeman effect according to the equation[7], $v_{1,2} = v_0 \pm \sqrt{E^2 + (g\mu_B B)^2}/h$. Here, $v_0 = D/h$ and $E/h$ are the ZFS parameters determined above. $g = 2$ stands for the Landé factor of $V_B^-$ in hBN, $\mu_B$ is the Bohr magneton and $h$ represents the Planck constant. This rather simple form will be addressed later in this paper, together with a discussion about the model. The spin-optical readout of Zeeman splitting in $V_B^-$ enables our polycrystalline bulk hBN sample to serve as a magnetic field quantum sensor.

Increasing the laser excitation power directly increases the PL brightness of the $V_B^-$ centers, thus improving the ODMR sensitivity. Therefore, we conducted power-dependent PL intensity measurements in Figure. 2e Typically, the power-dependent PL intensity of the $V_B^-$ follows this equation[8]:

$$I = \frac{I_{\text{sat}}}{\left(1 + \frac{P_{\text{sat}}}{P}\right)}, \qquad (1)$$

where $I$ is the PL intensity of the $V_B^-$, $I_{\text{sat}}$ is the saturation PL intensity, $P$ is the excitation laser power and $P_{\text{sat}}$ is the saturation laser power.

However, due to the high density of $V_B^-$ inside our bulk polycrystalline hBN, only linear dependent PL intensity can be observed in this material, which indicates the laser excitation power is far away from saturation. A maximum PL intensity of 2.2 Mcounts/s can be observed under 1.5 mW laser excitation.

Next, we investigated the effect of microwave power on the ODMR contrast by varying it from 1 mW to 630 mW while maintaining a constant laser excitation power of 350 μW. The ODMR contrast exhibited a two-order of magnitude increase, rising from 0.02% at 1 mW to 1.9% at 630 mW microwave power, reaching our setup limit. However, the near-linear increase tendency between the microwave power and absolute value of ODMR contrast indicates that the microwave power is still far from saturation (see Figure. 2f). The $V_B^-$ defects generated by He$^+$ irradiation predominantly reside on the front surface of the hBN sample within approximately 5 μm and remain distant from the MW co-planar waveguide. However, $B_{MW}$ propagates in the in-plane direction and decays significantly with increasing out-of-plane distance. In contrast, in hBN monocrystalline flakes directly transferred onto Au waveguide surfaces, the much smaller distance between $V_B^-$ defects and microwaves, combined with Au plasmonic enhancement, offers superior spin pumping performance under low-power microwave excitation[8]. This effect can be mitigated by dispersing irradiated hBN powder on the co-planar waveguide.

Next, we determined the sensitivity of $V_B^-$ in hBN bulk to external magnetic fields applied perpendicular to the sample surface (0°) under optimum laser excitation power (350 μW) and microwave power (630 mW). The sensitivity of $V_B^-$ can be deduced as follows[8]:

$$\eta_B \approx P_F \times \frac{h}{g\mu_B} \times \frac{\Delta v}{C\sqrt{R}}, \qquad (2)$$

where $P_F \approx 0.7$ is a numerical parameter for the Gaussian profile, $\Delta v = 110$ MHz is the linewidth of the ODMR signal under zero-magnetic field conditions, $C=1.9\%$ is the ODMR contrast and $R$ is the photon count rate. The PL intensity in photon count rate is measured by a photon counting system, giving a photon count rate of 516 kcounts/s under 350 μW laser excitation, resulting in a sensitivity of 200 μT/Hz$^{1/2}$, which is two orders of magnitude lower than that of Au plasmonic-enhanced hBN single crystal flakes[7].

The low sensitivity of $V_B^-$ in bulk hBN polycrystalline material can be attributed to the polycrystalline nature of the hot-pressed hBN, its relatively low PL intensity under laser excitation and low ODMR contrast. In contrast, hBN single crystal flakes directly transferred onto Au co-planar waveguides exhibit enhanced sensitivity. This enhancement is due to the coupling of free electrons on the metallic surfaces with electromagnetic waves, resulting in collective oscillations. These oscillations significantly reduce radiative decay times and substantially increase the PL brightness of the $V_B^-$ defect center by more than one order of magnitude[8].

Similarly, the relatively low ODMR contrast observed in our hBN samples results from the mixing of spin orientations within the bulk material. In hBN single crystal flakes, $V_B^-$ defects with spins oriented along the out-of-plane direction can align perpendicularly to the microwave magnetic driving field parallel to the waveguide surface, thereby maximizing ODMR contrast. However, in bulk polycrystalline hBN material, the random crystalline orientation between different grains leads to a varied orientation of $V_B^-$ spin defects relative to our microwave field. This randomness results in a reduced ODMR contrast.

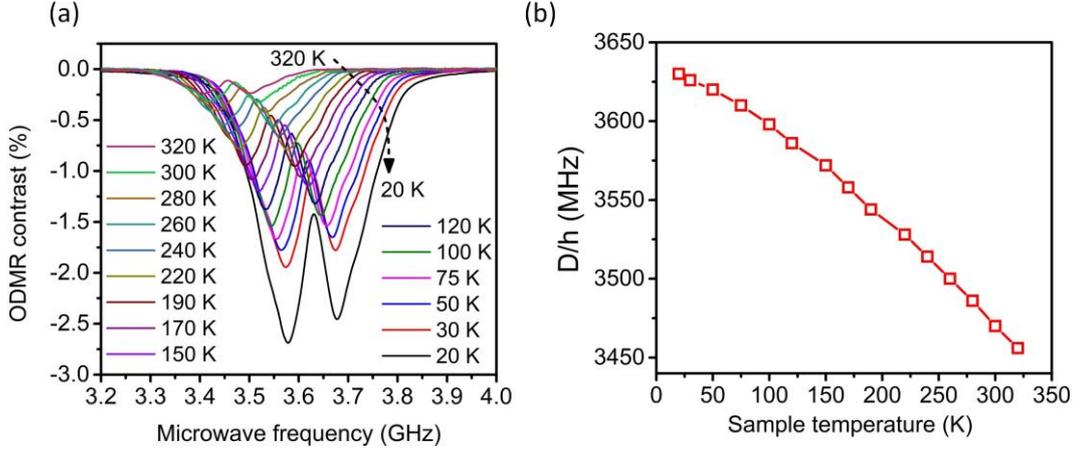

**Figure 3. Temperature-dependent ODMR measurements of polycrystalline hBN. a,** ODMR zero-field splitting of polycrystalline hBN under different temperatures. **b,** Temperature-dependent longitudinal ZFS parameter *D/h* and transverse ZFS parameter *E/h*.

For the zero-field splitting (ZFS) of VB- in hBN material, the longitudinal ZFS parameter D/h is highly dependent on the lattice parameters a and c of the hBN crystal, which shows high sensitivity to the lattice temperature and provides a good candidate for cryogenic thermometer. Figure. 3a shows the temperature-dependent ODMR signal without external magnetic field, and the microwave power is reduced to 50 mW to avoid the heating effect of the PCB waveguide. When the temperature is decreased from 320 K to 20 K, the ODMR resonance signal shows an almost linear shift toward higher frequency, which can be attributed to the lattice parameter difference under different temperature. Also, the ODMR signal contrast shows a strong increase for about 10 times, which also improves the ODMR sensitivity at low temperature. The temperature-dependence of the longitudinal ZFS parameter *D/h* (shown in Figure. 3b) can be attributed to the structural deformations of the hBN crystal lattice under different temperature and cause the difference of the spin-defect wave function delocalization in VB-. The resulting temperature-induced shift in $\Delta D(T)/h$ can be expressed as follows[19]:

$$\Delta D(T)/h = D(T)/h - D(300\,K)/h = \theta_a \eta_a + \theta_c \eta_c \qquad (4)$$

While $D(T)/h$ is the longitudinal ZFS parameter under temperature $T$, $D(300 K)/h$ is the longitudinal ZFS parameter under 300 K, factor $\theta_a=(-81\pm12)$ GHz, $\theta_b=(-24.5\pm0.8)$ GHz. $\eta_a$ and $\eta_c$ are the lattice parameters ($a$ and $c$) difference between $T$ and 300K[19]:

$$\eta_a = \frac{a(T)-a(300\ K)}{a(300\ K)}, \quad \eta_c = \frac{c(T)-c(300\ K)}{c(300\ K)} \tag{5}$$

Compared with the typical NV⁻ center in diamond[20], which shows only a small shift of $\Delta D/h = 7$ MHz over the temperature range of 10-295 K, our hBN sample shows a much more obvious and nearly linear shift of $\Delta D/h = 160$ MHz in the range of 20 K to 300 K (Figure. 3b), which makes our saleable hBN polycrystalline (sample size in mm scale) a nice candidate for in-suit sample temperature calibration and cryogenic thermometer application.

To demonstrate the alignment-free sensitivity of $V_B^-$ defects in hot-pressed polycrystalline hBN, we used a Helmholtz coil and measured the Zeeman splitting under three magnetic field orientations (X, Y, and Z) with a constant and homogeneous external magnetic field of 3.2 mT (see Figure. 4a).

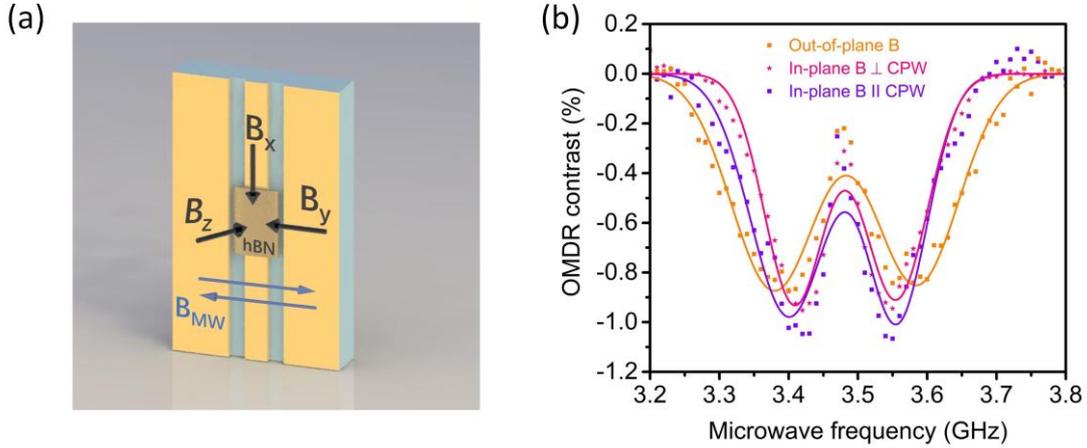

**Figure 4. Alignment-free magnetic field sensing with $V_B^-$ in hot-pressed hBN. a,** Schematic of the measurement procedure used to demonstrate the alignment-free of the magnetic field sensitivity in the hot-pressed hBN sample with $V_B^-$. $B_x$: in-plane field along the X-axis, parallel to the CPW and perpendicular to the microwave-induced magnetic field ($B_{MF}$), which lies in-plane along the Y-axis. $B_y$: in-plane external magnetic field along the Y axis, parallel to $B_{MF}$ and perpendicular to the coplanar waveguide (CPW). $B_z$: out-of-plane external magnetic field along the Z axis. **b,** ODMR resonance splitting under a 3.2 mT external magnetic field along each of the X, Y and Z directions.

As shown in Figure. 4b, under a 3.2 mT external magnetic field applied in three different orientations (X, Y, and Z axes), the ODMR spectrum of $V_B^-$ defects in polycrystalline hBN exhibits a Zeeman splitting and ODMR contrast across all magnetic field directions. In contrast, single NV⁻ centers (S = 1) in diamond, subjected to a similar external magnetic field strength, display a strong dependence on the magnetic field orientation, with Zeeman splitting decreasing by more than 95% as the off-axis angle between the spin quantization axis and the external magnetic field increases from 0 to 90 degrees[21]. This result validates the alignment-free of our hBN sample, confirming its potential as an all-directional magnetic field sensor.

Notably, we observed that under a constant external magnetic field, the Zeeman splitting is slightly smaller when the magnetic field is applied in the in-plane X and Y axes as compared to the out-of-plane Z axis. To further explore the potential directional dependence of the Zeeman splitting, we performed numerical simulations of ODMR spectra under magnetic fields applied along all three spatial directions. This anisotropic response can be reproduced by assuming a partial preferential alignment of the $V_B^-$ centers along the Z axis, likely arising from a mild out-of-plane texture in the polycrystalline hBN. A best match to the experimental ODMR spectra was achieved by introducing 30% more defects oriented along the Z axis in the simulation. While further experimental validation is needed, this result suggests that hot-pressed hBN may exhibit structural anisotropy due to grain alignment induced during fabrication. We describe the model we used to determine the resonance frequencies as follows.

The electron spin ground state Hamiltonian for a single $V_B^-$ defect reads (in units of $\hbar = 1$),

$$H = DS_z^2 + E(S_x^2 - S_y^2) + g\mu_B \vec{S} \cdot \vec{B}. \tag{6}$$

The first and second terms in Eq. (6) represent the longitudinal and transverse components of the ZFS, respectively. The last term describes the Zeeman interaction with an external magnetic field $\vec{B}$, where $\vec{S} = (S_x, S_y, S_z)$, denotes the spin vector operator, and $S_i$ are the spin-1 Pauli matrices, $i=x,y,z$. When the magnetic field is applied along the out-of-plane Z axis, the Hamiltonian can be diagonalized analytically, yielding eigenenergies $E_1 = D + \sqrt{E^2 + (g\mu_B B_z)^2}$, $E_2 = 0$, $E_3 = D - \sqrt{E^2 + (g\mu_B B_z)^2}$. In this configuration, the eigenstates remain the bare spin states $|+1\rangle, |0\rangle, |-1\rangle$. This alignment, combined with the preferential orientation of the spin quantization axes along the Z-direction, accounts for the resonance frequencies observed in Figure. 2d, which resemble those of an isolated spin-1 system. In the following section, we turn to numerical simulations to explore the system's behavior under different magnetic field orientations.

In polycrystalline hBN, the random orientation of grains leads to a distribution of defect quantization axes in all directions, making it impossible to define a unique quantization axis for each defect. Consequently, we adopt the laboratory frame, determined by the geometry of our ODMR setup, as the reference frame for defining the principal axes. In this frame, the microwave field is aligned along the X-axis and the $B_{MF}$ along the Y-axis, while the external magnetic field is applied sequentially along three orthogonal directions: Bx, By, and Bz, as illustrated in Figure. 4a.

To simulate the spin dynamics of the system under these conditions, we employed a Markovian Lindblad master equation (with $\hbar=1$) to capture both coherent evolution and decoherence processes [22, 23],

$$\frac{d\rho}{dt} = -i[H_t, \rho] + \frac{\Gamma}{2}(2S_x \rho S_x - S_x^2 \rho - \rho S_x^2), \tag{7}$$

where ρ is the density matrix, and $\Gamma = 1/T_1$ is a phenomenological decay rate associated with a longitudinal relaxation time $T_1 = 14$ μs [24]. The total Hamiltonian $H_t = H + g\mu_B B_{MF} \cos(\omega t) S_y$ includes the static spin Hamiltonian $H$ and a driving field oscillating at frequency ω along the Y-axis.

To model the polycrystalline nature of the sample, we numerically solved the time evolution of 1000 spin systems, each randomly oriented with respect to the lab frame, plus 300 systems oriented along Z (out-of-plane). The resulting time-dependent signal for each spin was averaged, and the spectrum was obtained using a Fast Fourier Transform (FFT).

Figure. 5a-c show good agreement between the experimental spectra and our model for the three orientations of the external magnetic field. Additionally, Figure. 5d presents the calculated Zeeman splitting of the resonance frequencies as a function of the magnetic field strength for the X, Y, and Z directions.

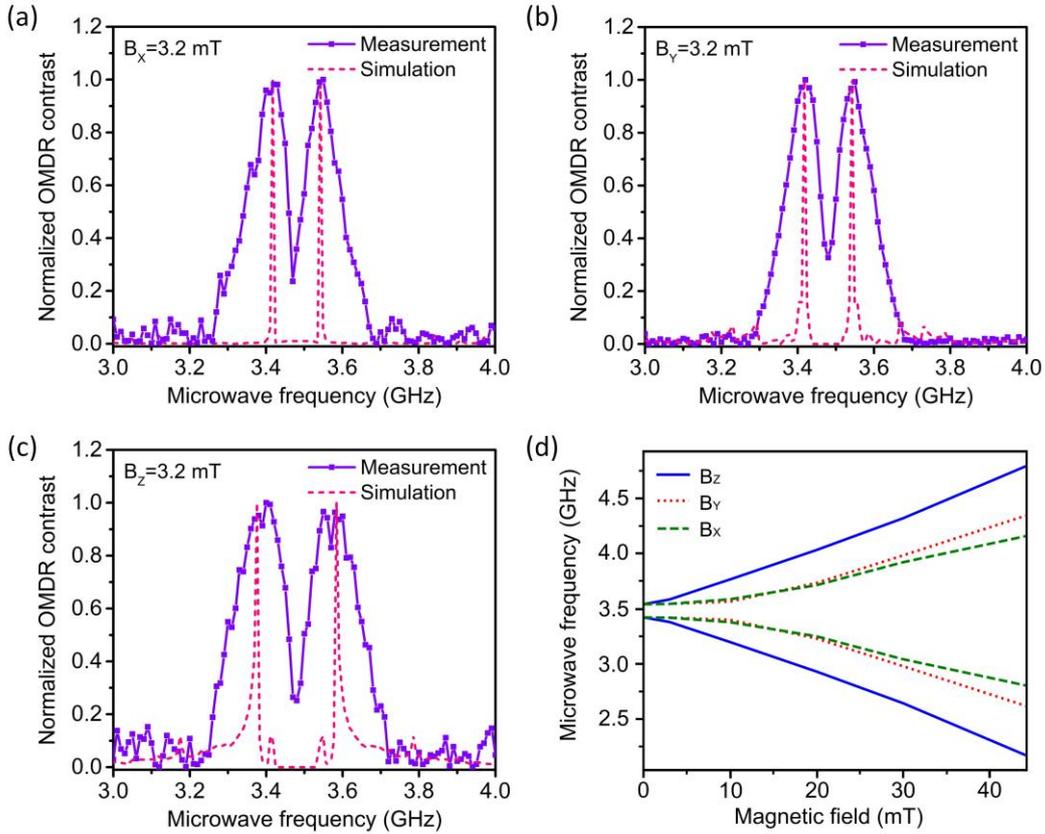

**Figure 5. Numerical simulation of alignment-free magnetic field sensing properties. a-c,** FFT of the averaged time-evolution simulated with Eq. (7) reproduces the spectrum for each magnetic field orientations. The amplitude was fixed to 3.2 mT. **d,** Resonance frequencies as a function of the magnetic field.

In summary, we have demonstrated room-temperature optically detected magnetic resonance from negatively charged boron vacancies in commercially available hot-pressed hBN substrates. The polycrystalline nature of the hot-pressed hBN, featuring a distribution of randomly oriented grains, enables magnetic field detection without the need for precise alignment between the defect's spin quantization axis and the external field. Although the sensitivity is not uniform across all directions, our numerical simulations indicate that vector magnetic field sensing is feasible through calibration, even in the presence of anisotropic defect orientation distributions. Compared to sensing platforms based on single-crystal hBN flakes, NV centers in diamond, or spin defects in silicon carbide, our approach offers a scalable, alignment-free, and mechanically robust alternative. These results establish hot-pressed hBN as a promising platform for the development of practical, low-cost quantum magnetic sensors suitable for applications in navigation, environmental monitoring, biomedical diagnostics, and beyond.

**Methods**

**Ion irradiation**

We conducted ion irradiation experiments at the Ion Beam Center of the Helmholtz-Zentrum Dresden-Rossendorf (HZDR) employing the 2 MV van de Graaf accelerator. The He ion beam energy was 1.7 MeV and a beam diameter of approximately 1 mm. The total He fluence provided was approximately $2\times10^{16}$ cm$^{-2}$, resulting in the generation of defects to a depth of approximately 5 µm.

**ODMR measurements**

We used a home-built confocal microscope to acquire ODMR spectra (Figure. 2b). We performed optical excitation by combining a 532 nm laser with various optical density (OD) filters to achieve the desired laser power output. We employed an objective lens (Olympus LCPLN50XIR 50x, NA=0.65) to focus the excitation laser beam. Subsequently, the PL signal was collected through the same objective lens and spectrally isolated from the laser light using a 550 nm dichroic mirror, followed by a long-pass 700 nm filter and a short-pass 950 nm filter. A silicon avalanche photodetector (Thorlabs, APD440A) detected the integrated PL signal from 700 nm to 950 nm (Figure. 1c).

We relied on synchronization and lock-in measurement techniques, generating a 22 Hz synchronization signal internally with the lock-in amplifier (Stanford Research SR830) to synchronize the entire system. The lock-in amplifier amplitude modulated the radiofrequency (RF) output using the synchronization signal, creating a periodic RF-on/RF-off state. During these synchronized RF-on/off states, we amplified the difference in PL (ΔPL) between events with and without resonant microwave excitation using the lock-in amplifier to achieve ΔPL readout.

To address the challenge of ΔPL amplitude being lower than the total PL level, we connected an electrical filter with a 1 dB cutoff of >1 Hz (Thorlabs, EF500) between the APD and the lock-in amplifier. This filter effectively blocked the direct component of the output current, primarily consisting of dark current from the detectors and PL signal from other optically active defects unresponsive to microwave excitation. Consequently, we collected only the alternating part of the output photocurrent originating from resonant VB$^-$ defects. This electrical filter allowed us to use a higher sensitivity circuit in the lock-in amplifier, resulting in a better signal-to-noise ratio.

# Reference

1. Li, W. & Wang, J. Magnetic Sensors for Navigation Applications: An Overview. *J. Navigation* **67**, 263–275 (2014).
2. Gerginov, V., Da Silva, F. C. S. & Howe, D. Prospects for magnetic field communications and location using quantum sensors. *Review of Scientific Instruments* **88**, 125005 (2017).
3. Aslam, N. *et al.* Quantum sensors for biomedical applications. *Nat Rev Phys* **5**, 157–169 (2023).
4. Yang, H., Sun, Y. & Wang, Y. Magnetic Sensing System for Potential Applications in Deep Earth Extremes for Long-Term Continuous Monitoring. *IEEE Trans. Instrum. Meas.* **71**, 1–11 (2022).
5. Gottscholl, A. *et al.* Spin defects in hBN as promising temperature, pressure and magnetic field quantum sensors. *Nat Commun* **12**, 4480 (2021).
6. Stern, H. L. *et al.* Room-temperature optically detected magnetic resonance of single defects in hexagonal boron nitride. *Nat Commun* **13**, 618 (2022).
7. Liang, H. *et al.* High Sensitivity Spin Defects in hBN Created by High-Energy He Beam Irradiation. *Advanced Optical Materials* **11**, 2201941 (2023).
8. Gao, X. *et al.* High-Contrast Plasmonic-Enhanced Shallow Spin Defects in Hexagonal Boron Nitride for Quantum Sensing. *Nano Lett.* **21**, 7708–7714 (2021).
9. Gao, X. *et al.* Nanotube spin defects for omnidirectional magnetic field sensing. *Nat Commun* **15**, 7697 (2024).
10. Rondin, L. *et al.* Magnetometry with nitrogen-vacancy defects in diamond. *Rep. Prog. Phys.* **77**, 056503 (2014).
11. Barry, J. F. *et al.* Sensitivity optimization for NV-diamond magnetometry. *Rev. Mod. Phys.* **92**, 015004 (2020).
12. Simin, D. *et al.* High-Precision Angle-Resolved Magnetometry with Uniaxial Quantum Centers in Silicon Carbide. *Phys. Rev. Applied* **4**, 014009 (2015).
13. Li, J. *et al.* Defect Engineering of Monoisotopic Hexagonal Boron Nitride Crystals *via* Neutron Transmutation Doping. *Chem. Mater.* **33**, 9231–9239 (2021).
14. Innocenzi, P. & Stagi, L. From Defects to Photoluminescence in h-BN 2D and 0D Nanostructures. *Acc. Mater. Res.* **5**, 413–425 (2024).
15. Haykal, A. *et al.* Decoherence of VB- spin defects in monoisotopic hexagonal boron nitride. *Nat Commun* **13**, 4347 (2022).
16. Udvarhelyi, P. *et al.* A planar defect spin sensor in a two-dimensional material susceptible to strain and electric fields. *npj Comput Mater* **9**, 150 (2023).
17. Baber, S. *et al.* Excited State Spectroscopy of Boron Vacancy Defects in Hexagonal Boron Nitride Using Time-Resolved Optically Detected Magnetic Resonance. *Nano Lett.* **22**, 461–467 (2022).
18. Liu, W. *et al.* Temperature-Dependent Energy-Level Shifts of Spin Defects in Hexagonal Boron Nitride. *ACS Photonics* **8**, 1889–1895 (2021).
19. Gottscholl, A. *et al.* Spin defects in hBN as promising temperature, pressure and magnetic field quantum sensors. *Nat Commun* **12**, 4480 (2021).
20. Chen, X.-D. *et al.* Temperature dependent energy level shifts of nitrogen-vacancy centers in diamond. *Applied Physics Letters* **99**, 161903 (2011).
21. Rondin, L. *et al.* Magnetometry with nitrogen-vacancy defects in diamond. *Rep. Prog. Phys.* **77**, 056503 (2014).

22. Breuer, H.-P. and Petruccione, F. *The Theory of Open Quantum Systems*. (OUP Oxford, 2022).
23. Johansson, J. R., Nation, P. D. & Nori, F. QuTiP: An open-source Python framework for the dynamics of open quantum systems. *Computer Physics Communications* **183**, 1760–1772 (2012).
24. Gao, X. *et al.* Nuclear spin polarization and control in hexagonal boron nitride. *Nat. Mater.* **21**, 1024–1028 (2022).

**Acknowledgments**
Support by the Ion Beam Center (IBC) at Helmholtz-Zentrum Dresden-Rossendorf (HZDR) is gratefully acknowledged. This work was partially funded by the Federal Ministry of Research, Technology and Space (BMFTR; project no. 31 46229 007) and the German Research Foundation (DFG; project no. 528206533).

**Author contributions**
Y.B. conceived the idea. S.W., P.W., S.P., and S.Z. prepared the samples and carried out the irradiation and optical characterizations. S.W. performed the ODMR measurements with support from Y.B. R.C. conducted the numerical modeling. Data analysis was carried out by S.W., R.C., and Y.B. The manuscript was written by S.W., and Y.B. with input from all authors. All authors discussed the results and contributed to the final version of the manuscript. M.H., J.W.L., S.Z., and Y.B. supervised and coordinated the entire project.

**Competing interests**
The authors declare no competing interests.

**Data availability**
The data that support the plots within this paper and other findings of this study are available from the corresponding authors upon request.